\documentclass[a4paper,fleqn]{cas-dc}
\usepackage[numbers,sort&compress]{natbib}
\usepackage[section]{placeins}
\usepackage{dcolumn}
\usepackage[utf8]{inputenc}
\usepackage[T1]{fontenc}
\usepackage{graphicx} 
\usepackage{float}

\usepackage[numbers]{natbib}
\usepackage{subfigure} 
\usepackage{caption}

\def\tsc#1{\csdef{#1}{\textsc{\lowercase{#1}}\xspace}}
\tsc{WGM}
\tsc{QE}
\tsc{EP}
\tsc{PMS}
\tsc{BEC}
\tsc{DE}

\begin{document}
\begin{sloppypar}

\title [mode = title]{On-Demand Pulse Shaping with Partially Coherent Pulses in Nonlinear Dispersive Media} 
                     
\author[1,2,3]{Qian Chen}

\author[1,2,3]{Yanlin Bai}

\author[1,2,3]{Xiaohan Wang} 

\author[4]{Peipei Peng\corref{cor1}} 
\cormark[1]
\ead{pengpeipei@szcu.edu.cn}

\author[1,2,3]{Jingsong Liu\corref{cor1}} 
\cormark[1]
\ead{liujs@sdnu.edu.cn}

\author[1,2,3]{Yangjian Cai\corref{cor1}} 
\cormark[1]
\ead{yangjiancai@sdnu.edu.cn}

\author[1,2,3]{Chunhao Liang\corref{cor1}} 
\cormark[1]
\ead{chunhaoliang@sdnu.edu.cn}

\address[1]{Shandong Provincial Engineering and Technical Center of Light Manipulations \& Shandong Provincial Key Laboratory of Optics and Photonic Device, School of Physics and Electronics, Shandong Normal University, Jinan 250014, China}
\address[2]{Collaborative Innovation Center of Light Manipulations and Applications, Shandong Normal University, Jinan 250358, China}
\address[3]{Joint Research Center of Light Manipulation Science and Photonic Integrated Chip of East China Normal University and Shandong Normal University, East China Normal University, Shanghai 200241, China}
\address[4]{School of Intelligent Manufacturing and Smart Transportation, Suzhou City University, Suzhou 215104, China}

\cortext[cor1]{Corresponding author}

\begin{abstract}
In this Letter, we employ the complex screen method to investigate the dynamic evolution of partially coherent pulses with specified properties as they propagate through a nonlinear Kerr medium. Our results reveal that partially coherent pulses can retain stable pulse characteristics and exhibit enhanced robustness when the source coherence is reduced. Importantly, by adjusting the source pulse properties, the far-zone pulse properties can be customized on demand, even in highly nonlinear environments. These findings are of significant importance for applications such as pulse shaping, free-space optical communication, information encryption etc. in nonlinear media. Notably, the results offer valuable insights for mitigating nonlinear effects in light beams within the spatial domain.
\end{abstract}

\maketitle

\section{Introduction}

Ultrashort pulses, characterized by extremely short durations and ultrahigh peak powers~\cite{1}, have found extensive applications in fields such as optical imaging, high-precision material processing, optical communication, quantum cryptography, and pump-probe techniques~\cite{2,3,4,5,6}. However, in practical scenarios, all natural pulses exhibit inherent fluctuations in amplitude, phase, or temporal duration, primarily due to spontaneous emission and imperfections in the laser sources, resulting in partial coherence. According to established optical coherence theory~\cite{7}, the second-order statistical properties of partially coherent pulses (PCPs) are described using the mutual coherence function in the time domain or the mutual spectral density in the frequency domain~\cite{8,9}. Unlike fully coherent pulses, PCPs introduce an additional degree of freedom$-$the degree of coherence (DOC)$-$which is a fundamental concept in optics. 

The “coherence-induced” effects observed in spatially partially coherent beams in free space can also be expected for PCPs in linear second-order dispersive media, owing to the well-established space-time analogy~\cite{10}. As a result, various PCPs with prescribed DOC have been proposed and studied, including cosine-Gaussian correlated Schell-model pulses, Laguerre-Gaussian correlated Schell-model pulses, multi-Gaussian correlated Schell-model pulses, and optical coherence lattices~\cite{11,12,13,14,15,16,17,18},among others. These pulses provide the flexibility to regulate pulse behavior during propagation and further shape the pulse as needed, facilitated by the DOC. To date, there are relatively few studies on the experimental generation of partially coherent pulses on demand, achieved through spectral manipulation techniques~\cite{15,18A}. Further, research on the interaction between PCPs and nonlinear dispersive media has primarily focused on Gaussian Schell$-$model pulses. Notably, studies have demonstrated that the statistical properties and probability density functions of PCPs during propagation can be controlled through nonlinear effects and the optical coherence of the source~\cite{19,20}. 

However, analyzing the propagation behavior of genuine PCPs in nonlinear dispersive media typically necessitates solving coupled nonlinear Schrödinger equations~\cite{9}, which presents substantial computational complexity. A straightforward strategy to alleviate this challenge is to represent a PCP as a superposition of coherent pulses, thereby reducing the problem to solving a single nonlinear Schrödinger equation. Several methods have been developed to achieve this, with the coherent mode (eigenmode) superposition method being the most widely used~\cite{7}. However, solving the homogeneous Fredholm integral equation to obtain the eigenmodes remains a formidable mathematical task~\cite{21}. As a result, only simple models, such as Gaussian Schell$-$model pulses, currently have corresponding solutions~\cite{7}, even though the method can be applied to any PCP. Notably, Ponomarenko et al.  ~\cite{22,23} demonstrated that this method can be extended to nonorthogonal pseudo$-$modes. Recently, we introduced complex and phase screen methods, adapting them from the spatial domain to the temporal domain~\cite{24}. With the aid of the Monte Carlo method, the random electric fields (viewed as random modes) of any Schell$-$model PCP can be efficiently obtained. This approach is both simple and powerful, providing significant advantages for studying coherence$-$related behaviors of PCPs during propagation in nonlinear media.

In this Letter, we employ the complex screen method~\cite{24} to obtain random electric fields of the customized PCPs. By applying the split-step Fourier method to the nonlinear Schrödinger equations, we systematically explore the evolution dynamics of the desired PCPs in the nonlinear Kerr medium, offering valuable insights into the interaction between nonlinearity and optical coherence.

\section{Theoretical model}

In the time domain, the second-order statistical properties of the PCPs with Schell$-$model type are described by the mutual coherence function~\cite{7}, given by:
\begin{equation}\label{eq1}
{\Gamma \left( {{t_1},{t_2}} \right) = \left\langle {{E^*}\left( {{t_1}} \right)E\left( {{t_2}} \right)} \right\rangle  = {\tau ^*}\left( {{t_1}} \right)\tau \left( {{t_2}} \right)\mu\left( {{t_2} - {t_1}} \right),}
\end{equation}
where the asterisk denotes the complex conjugate and $E(t)$ represents the instantaneous electric field of the stochastic pulses at time $t$. $\tau(t)$ is the complex amplitude. $\mu({{t_2} - {t_1}})$ stands for the DOC, which depends solely on the time difference between ${t_2}$ and ${t_1}$. By employing the complex screen method, the corresponding electric field is effectively represented as:
\begin{equation}\label{eq2}
	E\left( t \right) = \tau \left( t \right) \times {F_T}\left[ {\sqrt {p\left( f \right)} {C_n}\left( f \right)} \right],
\end{equation}
where ${F_T}$ denotes the Fourier transform and $p(f)$ is power spectral density function at frequency $f$, which can be attained through inverse Fourier transform of the DOC, i.e., $p = F_T^{ - 1}\left[ \mu  \right]$. ${C_n}\left( f \right)$ is a one-dimensional random complex function, generated by [randn(1, K)$ + i\times$randn(1, K)]$/$$\sqrt2$ in MATLAB, where K is the number of sampling points. Therefore, the corresponding electric field of any stochastic pulse with a Schell-model type can be determined if its complex amplitude $\tau \left( t \right)$ and power spectral density function $p(f)$ are known in advance. Further details could be found in Ref.~\cite{24}.

Once the instantaneous electric field is obtained, its evolution dynamics in the nonlinear Kerr medium are governed by the established nonlinear Schrödinger equation~\cite{24}, which is expressed as follows:
\begin{equation}\label{eq3}
i\frac{{\partial E}}{{\partial z}} - \frac{{{\beta _2}}}{2}\frac{{{\partial ^2}E}}{{\partial {t^2}}} =  - \gamma {\left| E \right|^2}E,
\end{equation}
where $E$ is the slowly-varying pulse envelope, ${\beta _2}$ and $\gamma $ are group velocity dispersion and Kerr nonlinearity coefficient, respectively. We have assumed that the time coordinate is defined in a reference frame traveling at the group velocity of the pulse. To explore generic features of the pulses, we rewrite the above equation in dimensionless form as follows, 
\begin{equation}\label{eq4}
i\sigma \frac{{\partial U}}{{\partial Z}} - {\mathop{\rm sgn}} \left( {{\beta _2}} \right)\frac{{{\sigma ^2}}}{2}\frac{{{\partial ^2}U}}{{\partial {T^2}}} =  - {\left| U \right|^2}U,
\end{equation}
here we define: $U = {E/\sqrt {\left\langle {P_0} \right\rangle } }$, $T = {t/t_p}$, $Z = z/L$, $L = \sqrt {{L_{NL}}{L_D}}$, and $\sigma  = \sqrt {{L_D}/{L_{NL}}}$, where $\left\langle {{P_0}} \right\rangle$ characterizes the average peak power, ${t_p}$ is the pulse with, defined by the distance between the two points on the intensity profile where the intensity value falls to 1/e of the peak intensity. ${L_{NL}} = {1/{\gamma \left\langle {{P_0}} \right\rangle }}$ as well as ${L_D} = {{t_p^2}/ {\left| {\beta _2} \right|}}$ stand for the typical nonlinear and dispersion lengths, respectively. $\sigma $ as the soliton parameter entirely determines the system dynamics, if the source state is given.

Next, we take the instantaneous electric field (in dimensionless form) of the stochastic pulses as the input for Eq.~(\ref{eq4}), and apply the split-step Fourier method to solve this equation. The output is considered as the instantaneous electric field ${U_1}\left( {t,z} \right)$ of the stochastic pulses at the receiving plane. We then refresh the random complex function ${C_n}\left( f \right)$ to update the input instantaneous electric field. By performing the same operation described above, we can obtain the new instantaneous electric field of the output pulse, ${U_2}\left( {t,z} \right)$. Through extensive iterative procedures, a large dataset of output instantaneous electric fields $\left\{ {{U_n}\left( {t,z} \right)} \right\}$ can be achieved. The intensity and the DOC of the stochastic pulses arriving at the receiving plane are approximately represented by

\begin{equation}\label{eq5}
I\left( {t,z} \right) \approx \sum\nolimits_N {{{{{\left| {{U_n}\left( {t,z} \right)} \right|}^2}} \mathord{\left/
			{\vphantom {{{{\left| {{U_n}\left( {t,z} \right)} \right|}^2}} N}} \right.
			\kern-\nulldelimiterspace} N}} ,
\end{equation}
and
\begin{equation}\label{eq6}
\mu \left( {{t_1},{t_2};z} \right) \approx \frac{{\sum\nolimits_N {U_n^*\left( {{t_1},z} \right){U_n}\left( {{t_2},z} \right)} }}{{\sqrt {\sum\nolimits_N {{{\left| {{U_n}\left( {{t_1},z} \right)} \right|}^2}} } \sqrt {\sum\nolimits_N {{{\left| {{U_n}\left( {{t_2},z} \right)} \right|}^2}} } }},
\end{equation}
respectively. $N$ is the total number of the output electric fields, and it should take a sufficiently large number to ensure that Eq.~(\ref{eq5}) and ~(\ref{eq6}) provide accurate approximations.

 Now we consider a complex PCP — a multi-Gaussian Schell-model pulse with a cosh-Gaussian intensity profile, referred to as a CG-MGSMP. Its complex amplitude $\tau '\left( T \right)$~\cite{25} and power spectral density function $p'\left( \upsilon  \right)$~\cite{17}, in dimensionless form in Eq.~(\ref{eq2}), are given by
 \begin{equation}\label{eq7}
\tau '\left( T \right) = {A_0}\cosh \left( {{\Omega _0}T} \right)\exp \left( { - \frac{1}{2}{T^2}} \right),
\end{equation}
and
\begin{equation}\label{eq8}
p'\left( \upsilon  \right) = \frac{{\sqrt {2\pi } {t_q}}}{{{C_0}}}\sum\limits_{m = 1}^M {\left(\begin{array}{l}
		M\\
		m
	\end{array}\right)} {\left( { - 1} \right)^{m - 1}}\exp \left( { - 2m{\pi ^2}t_q^2{\upsilon ^2}} \right),
\end{equation}
respectively, where the constant $A_0$ is used to ensure that the pulse's peak power is unity. The parameter ${\Omega _0}$ associated with cos-hyperbolic part governs the degree of decentralization. The coefficient ${C_0}$ is given by ${C_0} = \sum\limits_{m = 1}^M {\frac{{{{( - 1)}^{m - 1}}}}{{\sqrt m }}\left( \begin{array}{l}
		M\\
		m
	\end{array} \right)}$ and $M$ is the mode number. Further, $t_q = t_c/t_p$ denotes the global coherence time and $\upsilon  = {t_p}f$ is the dimensionless frequency. By substituting Eq.~(\ref{eq2}), (\ref{eq7}) and (\ref{eq8}) into the dimensionless nonlinear Schrödinger equations in Eq.~(\ref{eq4}), we can investigate the pulse behavior of the CG-MGSMP propagating in the nonlinear Kerr medium using the split-step Fourier method. Unless stated otherwise, the relevant parameters are set as $\Omega_0$ = 2, $M$ = 40, $t_p$ = 10 ps, $\beta_2$ = 20 ps$^2$/km, N=50 000 and $\gamma$ = 0.1w$^{-1}$km$^{-1}$. Furthermore, we vary the soliton parameter $\sigma$ by adjusting the average peak power $\left\langle {{P_0}} \right\rangle$.
  
\section{Simulation results and analysis}

\begin{figure}[h]
	\centering
	\includegraphics[width=1\linewidth]{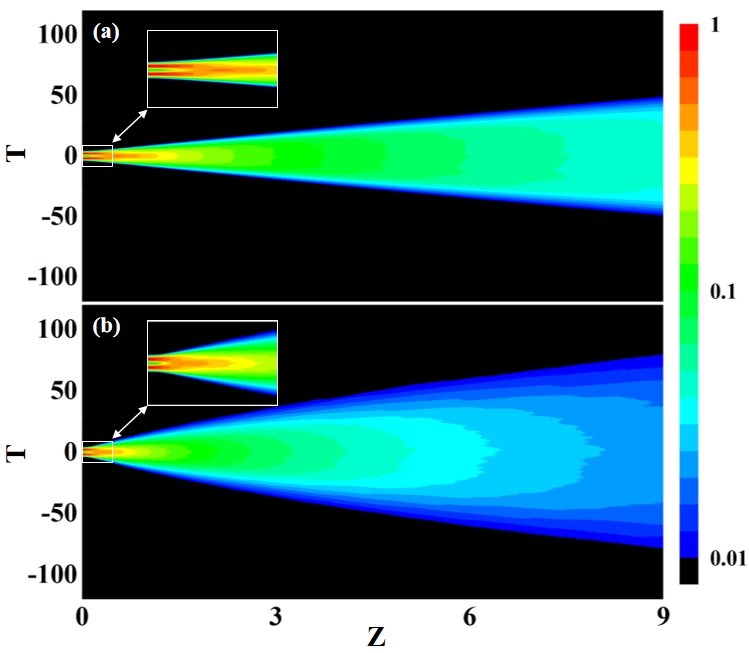}
	\caption{Evolution of the normalized intensity (${I\left( T \right)/ \left\langle {{P_0}} \right\rangle }$) of the CG-MGSMPs during propagation in the (a) linear ($\sigma=0$) and (b) nonlinear ($\sigma=5$) dispersive media. The global coherence time is set to ${t_q} = 0.6$.}
	\label{figure1}
\end{figure}

\begin{figure*}[h]
	\centering
	\includegraphics[width=0.8\linewidth]{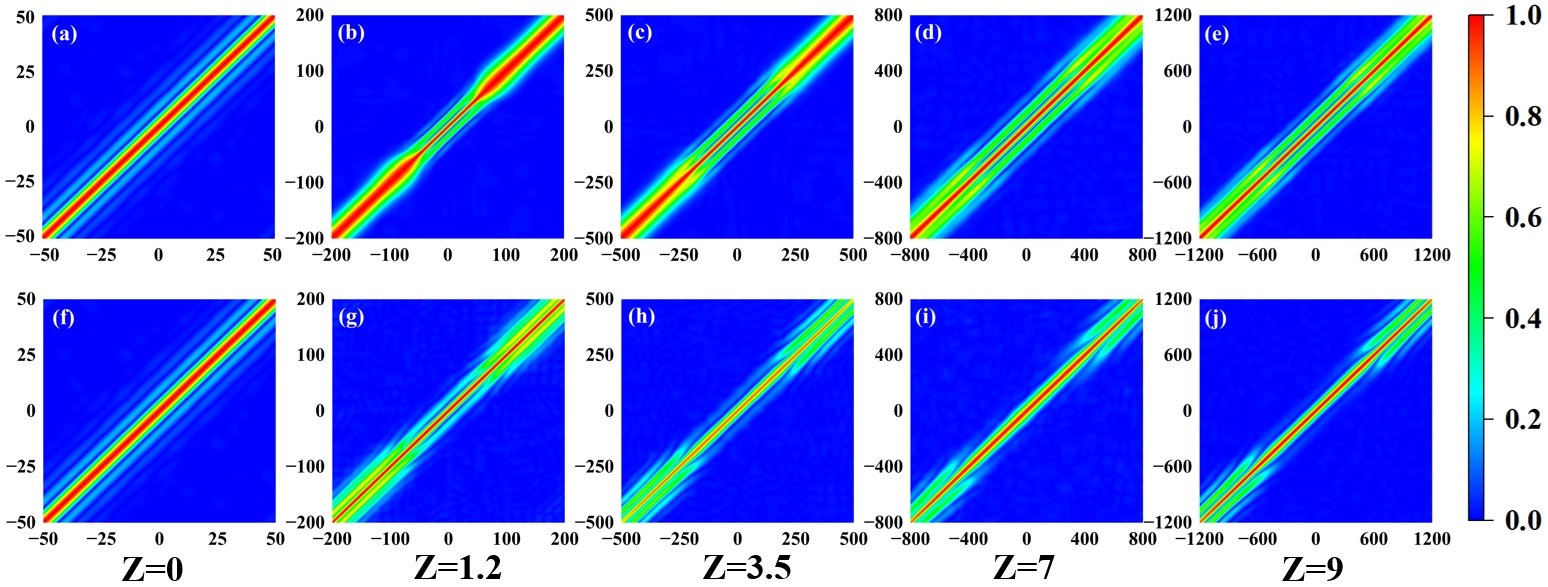}
	\caption{Evolution of the modulus of DOC ($\left| {\mu \left( {{T_1},{T_2}} \right)} \right|$) of the CG-MGSMPs during propagation in the linear ($\sigma=0$) and nonlinear ($\sigma=5$) dispersive media, shown in the top and bottom rows, respectively. The global coherence time is set to ${t_q} = 0.6$.}
	\label{figure2}
\end{figure*}

In this section, we explore the evolution dynamics of the CG-MGSMPs during propagation. To highlight the effect of nonlinearity, we also present the results of the linear case for comparison, introducing the dimensionless propagation distance $Z = {z/{L_D}}$. First, we present density maps illustrating the entire process of normalized intensity evolution, shown in Fig.~\ref{figure1}. In the linear case [Fig.~\ref{figure1}(a)], the pulse intensity, initially represented by a double Gaussian at the light source plane (determined by the complex amplitude $\tau '\left( T \right)$), gradually converges during transmission, resulting in a flat-top pulse distribution in the far field. This process is primarily governed by the source DOC, as discussed in more detail below. In contrast, under nonlinear effects [Fig.~\ref{figure1}(b)], the pulse initially experiences intensified convergence followed by significant divergence. As a result, the pulse evolves into a Gaussian-like profile in the far field, as further explained below. For the modulus of the DOC evolution of CG-MGSMPs during propagation in both linear and nonlinear dispersive media, we show the relevant results in the top and bottom rows of Fig.~\ref{figure2}, respectively. In the linear case, the DOC gradually transitions from a Schell-model type at the source plane to a non-Schell-model type, i.e., $\mu \left( {{T_1},{T_2}} \right) \ne \mu \left( {{T_2} - {T_1}} \right)$, during transmission. This transition is the result of the combined effects of the pulse source intensity and DOC. As the propagation distance increases, the shape of the degree of coherence no longer changes; rather, its size scales linearly. This phenomenon is also observed in nonlinear media. However, nonlinear effects can induce decoherence and alter the distribution of the DOC, as seen in the comparison between Fig.~\ref{figure2}(e) and Fig.~\ref{figure2}(k). Thus, nonlinearity introduces distortion in the CG-MGSMP, a conclusion that applies to any PCP, and is well-established for fully coherent pulses. 
	
\begin{figure*}[h]
	\centering
	\includegraphics[width=0.8\linewidth]{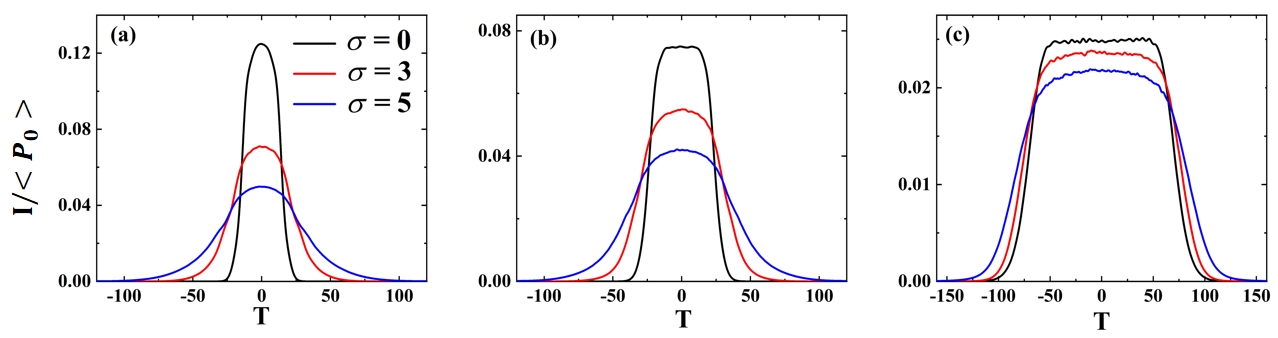}
	\caption{The normalized intensity (${I\left( T \right)/ \left\langle {{P_0}} \right\rangle }$) distributions of the CG-MGSMPs for different values of the soliton parameter $\sigma$ and global coherence time ${t_q}$ at the propagation distance $Z=5$. From left to right, the global coherence time is set to 1, 0.6, and 0.2, respectively. }
	\label{figure3}
\end{figure*}

\begin{figure}[h]
	\centering
	\includegraphics[width=\linewidth]{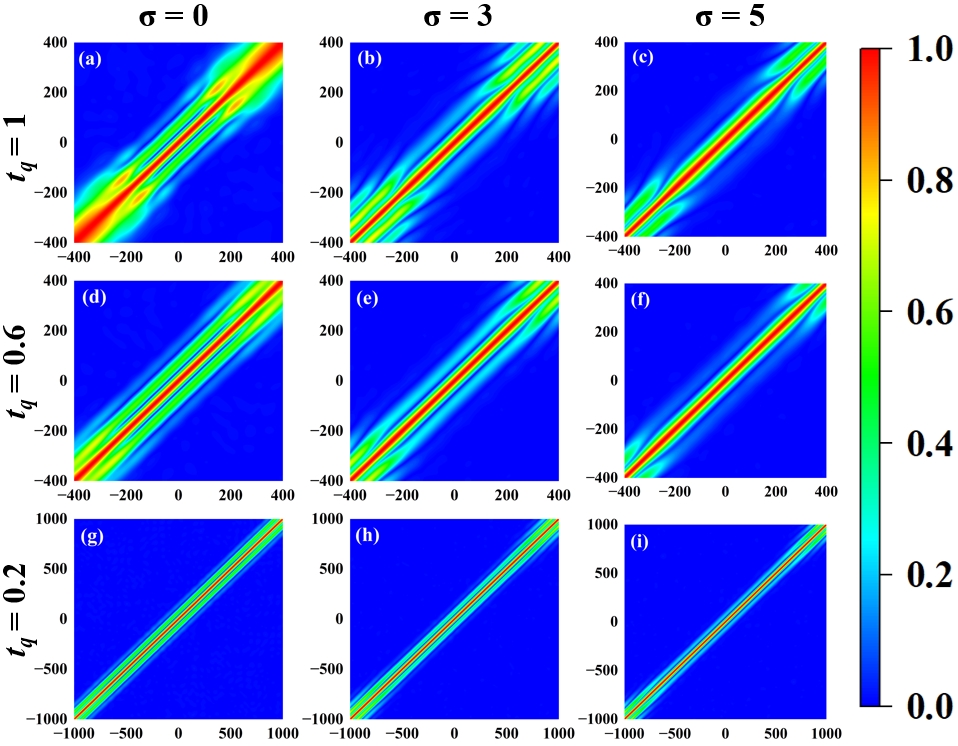}
	\caption{The modulus of DOC $\left| {\mu \left( {{T_1},{T_2}} \right)} \right|$ distributions of the CG-MGSMPs for different values of the soliton parameter $\sigma$ and global coherence time ${t_q}$ at the propagation distance $Z=5$.}
	\label{figure4}
\end{figure}

Next, we study the impact of the soliton parameter $\sigma $ (which determines the nonlinearity) and global coherence time ${t_q}$ on pulse properties at the receiving plane. The propagation distance is set to $Z=5$. The normalized intensity curves are plotted in Fig.~\ref{figure3}. For high coherence [see Fig.~\ref{figure3}(a)], the pulse intensity maintains a Gaussian-like profile, as it is jointly determined by the source intensity and DOC. However, as the nonlinearity increases, the pulse undergoes divergence, as discussed earlier. When the global coherence time is reduced to ${t_q} = 0.6$ [see Fig.~\ref{figure3}(b)], the pulse exhibits a flat-top distribution in the linear case, with the source DOC being the dominant factor. For higher nonlinearity (larger value of $\sigma$), the flat-top intensity profile shifts back to a Gaussian-like distribution. For low coherence [see Fig.~\ref{figure3}(c)], one finds that the pulses exhibit the a flat-top profile, which is almost independent of $\sigma$.  It can be inferred that the CG-MGSMPs with low coherence are robust against the nonlinearity. More important, we are able to shape the pulse intensity on demand through the source DOC, even under high nonlinearity. In Fig.~\ref{figure4}, we present the modulus of DOC distributions of the CG-MGSMPs at propagation distance $Z=5$. For high coherence (see first row), as nonlinear effects intensify, the DOC undergoes significant changes, accompanied by pronounced decoherence phenomena. With global coherence time ${t_q}$ decreasing, the DOC transitions from a non-Schell-mode type to a Schell-model type in both linear and nonlinear cases. According to coherence statistics theory~\cite{7}, the DOC distribution for low-coherence pulse is determined solely by the source intensity, and the two are approximately related by a Fourier transform. Notably, for the low coherence case ${t_q} = 0.2$ (see last row), the DOC distribution is almost unaffected by nonlinear effects, demonstrating strong robustness. Of course, it can be predicted that as the coherence time decreases further, the robustness of PCPs will be enhanced, which is of significant importance for pulse manipulation in nonlinear dispersive media.

To further quantitatively assess the impact of source optical coherence on mitigating nonlinear effects, we use a similarity function to quantitatively evaluate the variations in both intensity and DOC of the CG-MGSMPs for different value of the soliton parameter $\sigma$ and global coherence time ${t_q}$. The similarity function is defined as 
\begin{equation}\label{eq9}
{S_\sigma } = \frac{{{{\left[ {\int {{\Omega _\sigma }\left( T \right){\Omega _0}\left( T \right)dT} } \right]}^2}}}{{\int {\Omega _\sigma ^2\left( T \right)dt} \int {\Omega _0^2\left( T \right)dT} }},
\end{equation}
where ${\Omega _\sigma }$ and ${\Omega _0}$ represent the normalized intensity or modulus of the DOC of the CG-MGSMPs at the receiving plane in nonlinear and linear dispersive media, respectively. 

\begin{figure}[h]
	\centering
	\includegraphics[width=1\linewidth]{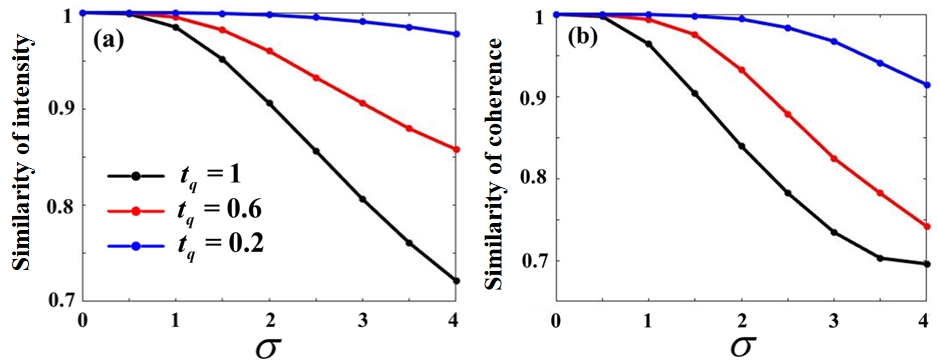}
	\caption{Similarity of the normalized intensity (a) and the modulus of the DOC (b) of the CG-MGSMPs at the propagation distance $Z=5$ as a function of the soliton parameter $\sigma$.}
	\label{figure5}
\end{figure}

We present the similarity curves for the normalized intensity and the modulus of the DOC of the pulse at the propagation distance $Z=5$, as a function of the soliton parameter $\sigma$, shown in the left and right panels of Fig.~\ref{figure5}, respectively. Both curves exhibit a decreasing trend with increasing soliton parameter $\sigma$. As the nonlinear effects intensify, the CG-MGSMPs inevitably undergo distortion, as discussed previously. However, it is noteworthy that for lower coherence values, the similarity of both intensity and DOC significantly improves. Even under high nonlinearity ($\sigma  = 4$), the similarity values for the intensity and DOC can reach approximately 0.98 and 0.92, respectively, at low coherence ${t_q} = 0.2$ Therefore, Fig.~\ref{figure5}, through the quantitative similarity analysis, clearly demonstrates that reduced optical coherence enhances robustness against nonlinear distortion. The results obtained for the CG-MGSMPs in this Letter are also applicable to any PCPs.

\section{Conclusion}
In this Letter, we employ the complex screen method to investigate the propagation behavior of the PCPs in nonlinear dispersive media. We demonstrate that, similar to fully coherent pulses, PCPs experience distortion and decoherence under the influence of nonlinear effects. However, when the source optical coherence of the stochastic pulse is reduced, the pulse properties are preserved, showing strong resistance to these nonlinear effects. Additionally, the far-zone pulse intensity is determined solely by the source DOC, while the far-zone DOC is determined solely by the source intensity. These properties follow a Fourier transform relationship, allowing us to tailor the pulse characteristics on demand in nonlinear dispersive media. Notably, in the spatial domain, we have advanced methods for high-capacity, high-fidelity, and high-security information encryption and optical imaging that are resilient to atmospheric turbulence, through the DOC ~\cite{26,27,28}. As discussed, the space-time analogy further implies significant potential for applications in information encryption and transmission in the temporal domain, expanding beyond pulse shaping in nonlinear media.
 \hspace*{\fill} \ 
 
\section*{CRediT authorship contribution statement}
\textbf{Qian Chen:} Writing – original draft, Validation, Methodology, Conceptualization; \textbf{Yanlin Bai:} Validation, Software, Formal analysis; \textbf{Xiaohan Wang:} Methodology, Investigation, Formal analysis; \textbf{Peipei Peng:} Validation, Methodology, Resources, Formal analysis; \textbf{Jingsong Liu:} Resources, Funding acquisition, Writing – review \& editing; \textbf{Yangjian Cai:} Resources, Funding acquisition, Writing – review \& editing; \textbf{Chunhao Liang:} Resources, Funding acquisition, Writing – review \& editing, Conceptualization.

\section*{Declaration of Competing Interest}
The authors declare that they have no known competing financial interests or personal relationships that could have appeared to influence the work reported in this paper. 

\section*{Data availability}
Data will be made available on request.

\section*{Acknowledgement}
This work was supported by the National Key Research and Development Program of China (2022YFA1404800), the National Natural Science Foundation of China (12374311, 12192254, 92250304, and W2441005), the Taishan Scholars Program of Shandong Province (tsqn202312163), the Qingchuang Science and Technology Plan of Shandong Province (2022KJ246), the China Postdoctoral Science Foundation (2022T150392), the Natural Science Foundation of Shandong Province (ZR2023YQ006 and ZR2024QA216), and the Natural Sciences and Engineering Research Council of Canada (RGPIN-2018-05497).



\end{sloppypar}

\begin{thebibliography}{99}
\bibitem{1}
J.-C. Diels and W. Rudolph, Ultrashort Laser Pulse Phenomena (Academic Press, San Diego, 1996).
\bibitem{2}
Hentschel, M., Kienberger, R., Spielmann, C. et al. Attosecond metrology. Nature 414, 509–513 (2001).
\bibitem{3}
Yin, J., Tan, Z., Hong, H. et al. Ultrafast and highly sensitive infrared photodetectors based on two-dimensional oxyselenide crystals. Nat. Commun 9, 3311 (2018). 
\bibitem{4}
Kraus, P.M., Zürch, M., Cushing, S.K. et al. The ultrafast X-ray spectroscopic revolution in chemical dynamics. Nat. Rev. Chem 2, 82–94 (2018).
\bibitem{5}
Shirk M D , Molian P A . A review of ultrashort pulsed laser ablation of materials. J. Laser Appl. 10,18-28 (1998). 
\bibitem{6}
Gisin N, Ribordy G, Tittel W, et al. Quantum cryptography. Rev. Mod. Phys.,  74,145,(2002). 
\bibitem{7}
L. Mandel and E. Wolf, Optical Coherence and Quantum Optics (Cambridge University Press, 1995)
\bibitem{8}
Pääkkönen P, Turunen J, Vahimaa P, et al. Partially coherent Gaussian pulses. Opt. Commun. 204, 53-58 (2002). 
\bibitem{9}
Lin Q, Wang L, Zhu S. Partially coherent light pulse and its propagation .Opt. Commun. 219, 65-70 (2003). 
\bibitem{10}
Lancis J, Torres-Company V, Silvestre E, et al. Space–time analogy for partially coherent plane-wave-type pulses. Opt. Lett. 30, 2973-2975 (2001). 
\bibitem{11}
Ma L, Ponomarenko S A. Optical coherence gratings and lattices. Opt. Lett. 39, 6656-6659 (2014). 
\bibitem{12}
Ding C, Koivurova M, Turunen J, et al. Temporal self-splitting of optical pulses. Phys. Rev. A 97, 053838(2018). 
\bibitem{13}
Koivurova M, Laatikainen J, Friberg A T. Nonstationary optics: tutorial. J. Opt. Soc. Am. A 41, 615-630 (2024).
\bibitem{14}
Liu H, Du Z, Li Y, et al. Self-focusing and self-splitting properties of partially coherent temporal pulses propagating in dispersive media. Opt. Express 31, 7336-7350 (2023). 
\bibitem{15}
Talukder R, Halder A, Koivurova M, et al. Generation of pulse trains with nonconventional temporal correlation properties. J. Opt. 24, 055502(2022). 
\bibitem{16}
Ding C, Korotkova O, Zhao D, et al. Propagation of temporal coherence gratings in dispersive medium with a chirper. Opt. Express 28, 7463-7474 (2020).
\bibitem{17}
Ding C, Korotkova O, Pan L. The control of pulse profiles with tunable temporal coherence. Phys. Lett. A 378, 1687-1690 (2014). 
\bibitem{18}
Ding C, Korotkova O, Zhang Y, et al. Cosine-Gaussian correlated Schell-model pulsed beam. Opt. Express 22, 931-942 (2014). 
\bibitem{18A}
Audo F, Rigneault H, Finot C. Linear and nonlinear fiber propagation of partially coherent fields exhibiting temporal correlations. Ann. Phys. 532, 1900597 (2020).
\bibitem{19}
Liang C, Ponomarenko S A, Wang F, et al. Rogue waves, self-similar statistics, and self-similar intermediate asymptotics. Phys. Rev. A 100, 063804(2019). 
\bibitem{20}
Liang C, Ponomarenko S A, Wang F, et al. Temporal boundary solitons and extreme superthermal light statistics. Phys. Rev. Lett. 127, 053901(2021). 
\bibitem{21}
Morse P M C, Feshbach H. Methods of theoretical physics[M]. Technology Press, 1946.
\bibitem{22}
Yang S, Ponomarenko S A, Chen Z. Coherent pseudo-mode decomposition of a new partially coherent source class. Opt. Lett. 40, 3081-3084 (2015). 
\bibitem{23}
Ponomarenko S A. Complex Gaussian representation of statistical pulses. Opt. Express 19, 17086 -17091 (2011). 
\bibitem{24}
Wang X, Tang J, Wang Y, et al. Complex and phase screen methods for studying arbitrary genuine Schell-model partially coherent pulses in nonlinear media. Opt. Express 30, 24222-24231 (2022). 
\bibitem{25}
Sharma V, Thakur V. Exploring the potential of cosh-Gaussian pulses for electron acceleration in magnetized plasma. J Opt., 1-7 (2024). 
\bibitem{26}
Peng, D., Huang, Z., Liu, Y. et al. Optical coherence encryption with structured random light. PhotoniX 2, 6 (2021). 
\bibitem{27}
Liu Y, Chen Y, Wang F, et al. Robust far-field imaging by spatial coherence engineering. Opto-Electron. Adv. 4, 210027(2021).
\bibitem{28}
Liu X, Ponomarenko S A, Wang F, et al. Incoherent mode division multiplexing for high-security information encryption. arXiv preprint arXiv:2304.06455, (2023).
\end{thebibliography}
\end{document}